\documentclass[aps,prl,twocolumn,groupedaddress,showpacs]{revtex4}

\usepackage{graphicx}

\begin{document}

\title{Melting of the Na layers in solid Na$_{0.8}$CoO$_2$}
\author{M. Weller}
\email[]{weller@phys.ethz.ch}
\homepage[]{www.solid.phys.ethz.ch/ott}
\author{A. Sacchetti, H. R. Ott, K. Mattenberger, B. Batlogg}
\affiliation{Laboratorium f\"ur Festk\"orperphysik, ETH Z\"urich, 8093 Z\"urich, Switzerland}

\date{\today}

\begin{abstract}
Data of $^{23}$Na NMR spectra- and relaxation measurements are interpreted as suggesting that,  upon increasing temperature the Na layers in Na$_{0.8}$CoO$_2$ adopt a 2D liquid state at $T=291$~K. The corresponding first order phase transition is preceded by a rapidly increasing mobility and diffusion of Na ions above 200~K. Above 291~K the $^{23}$Na NMR response is similar to that previously observed in superionic conductors with planar Na layers.
\end{abstract}

\maketitle

The recent activities in probing the physical behaviour of the compound series Na$_x$CoO$_2$ ($0.3<x<1$) revealed a number of exciting features with respect to structural, electronic, magnetic and superconducting properties upon varying $x$~\cite{Takada2003, Huang2004a, Pedrini2005, Roger2007, Schulze2008, deVaulx2005, Julien2008, Alloul2008}. Quite generally these compounds may be regarded as composed of stacks of alternating Na and Co-O layers. The role of the Na layers and the $x$ dependence of the Na-ion arrangements have been discussed in detail~\cite{Roger2007,Zhang2005}. The itinerant electronic and the magnetic degrees of freedom are concentrated in the Co-O layers~\cite{Gavilano2006}. Depending on the value of the Na concentration $x$, different ground states were identified. In this work we report a detailed $^{23}$Na NMR investigation of the compound Na$_{0.8}$CoO$_2$. As we outline below our data are to some extent in conflict with previously published details of structural features of the Na sublattice, in particular if $x<1$~\cite{Roger2007}. Results of diffraction experiments were discussed in relation with models describing selected static vacancy arrangements~\cite{Roger2007,Zhang2005}; aspects of the motion of Na-ions have been addressed to a much lesser extent~\cite{Schulze2008,Gavilano2005}. Our results give clear evidence that the dynamics of the Na ions is a dominating factor for the understanding of these compounds at temperatures exceeding 200~K where a growing diffusion of the Na-ions sets in. Upon increasing temperature, this motion leads to a first order phase transition, most likely indicating the melting of the 2D layers of the Na sublattice at 291~K.

\begin{figure}
\includegraphics[width=\columnwidth]{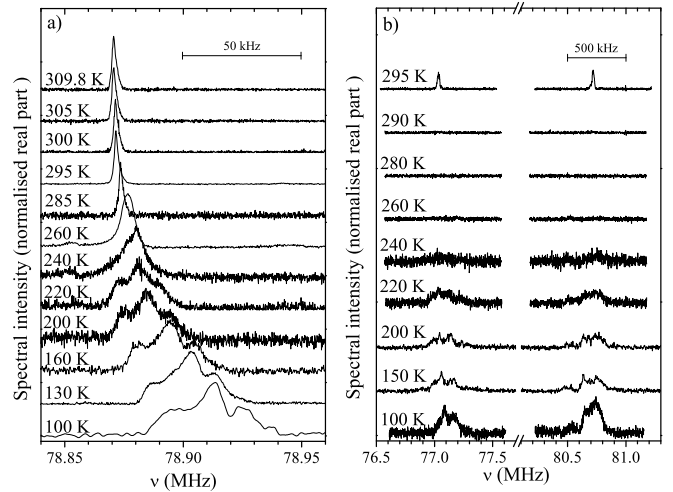}
\caption{Normalised $^{23}$Na NMR spectra of Na$_{0.8}$CoO$_2$ at different temperatures recorded in an external magnetic field of $B_0=7$~T. a) Central line; b) Quadrupolar satellites measured on both sides of the central line. Their amplitudes are multiplied by a factor 5 with respect to the signals in plot a). \label{Spectra}}
\end{figure}

The $^{23}$Na nuclear magnetic resonance (NMR) experiments probed a 10.3~mg single crystalline specimen of Na$_x$CoO$_2$ ($x=0.80(1)$). Results of X-ray diffraction experiments confirmed the high structural quality of the sample and a quantitative chemical analysis of a specimen chosen from the same batch confirmed the Na concentration $x$. The crystal was characterised by magnetic susceptibility measurements using a commercial SQUID magnetometer. The NMR experiments were made in a commercial $^4$He gas-flow cryostat, and the fixed external magnetic field $B_0=7$~T was provided by a commercial, ultra-shielded magnet. The $^{23}$Na NMR spectra of the central line and the two quadrupolar satellites were separately obtained by fast Fourier transforming (FFT) the second half of the corresponding spin echo signals~\footnote{The employed spin echo pulse sequence for all the spectra shown in fig.~\ref{Spectra} is 1~$\rm \mu s$, 30~$\rm \mu s$, and 2~$\rm \mu s$ for the first pulse, the interpulse delay, and the second pulse, respectively.}. For measuring the spin-lattice relaxation rate, a pulse sequence with a single preparation pulse was imposed, followed by a variable waiting time and a subsequent spin-echo experiment. The spin-spin relaxation rate (SSRR) was obtained by monitoring spin echoes after varying interpulse delay times.

In figure~\ref{Spectra} we show a selection of $^{23}$Na NMR spectra of the central line and the quadrupolar satellites in panels a) and b), respectively. While we note a drastic change, i.e., an abrupt quenching of the quadrupolar satellite component of the spectrum between 295 and 290~K, we observe no significant variation of the rather narrow and Lorentzian shaped central line in the same temperature interval. Upon further reduction of the temperature, the quadrupolar satellites reappear below $270$~K~\footnote{Although the quadrupolar satellite is hardly visible in the spectrum at $T=260$~K, we verified that spin-echo signals are detectable up to almost 270~K}. The frequency splitting between the satellites at $T<270~K$ is almost the same as that observed at 292~K and above. Below 270~K, the central line broadens and splits up into at least 3 lines, indicating inequivalent sites for the Na ions, and the centre of gravity of the signal shifts to higher frequencies. Concomitantly the satellite signal has broadened, exhibits a more complicated shape, but its width remains approximately constant with decreasing temperature.

The spin-lattice relaxation rate (SLRR) $T_1^{-1}$ and the spin-spin relaxation rate (SSRR) $T_2^{-1}$, both as a function of temperature between 100 and 310~K, were obtained from appropriate fits to the measured magnetisation recovery curves. It turned out that different relaxation channels have to be considered for the evaluation of $T_1^{-1}(T)$ and we refer to the work of Suter et al.~\cite{Suter1998} and refs.~therein. It outlines how to calculate transition probabilities between nuclear-spin energy levels for various microscopic relaxation mechanisms and how to calculate the time evolution of the magnetisation recovery for various irradiation conditions.
\begin{figure}
\includegraphics[width=\columnwidth]{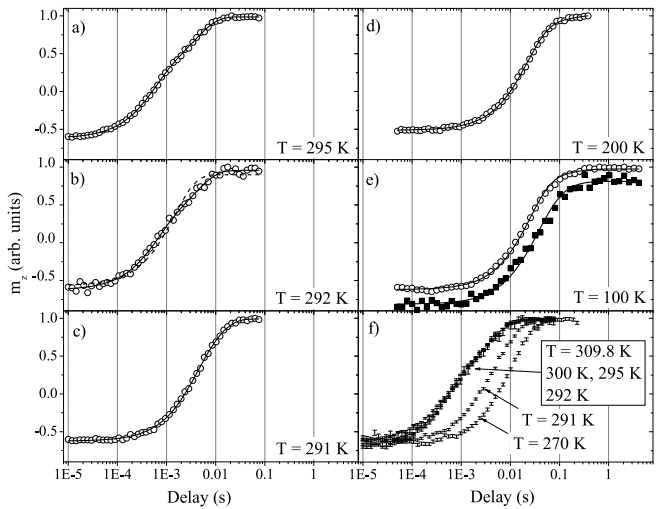}
\caption{$^{23}$Na NMR magnetisation recovery curves of Na$_{0.8}$CoO$_2$ recorded at selected temperatures after irradiation of the central line ($-\frac12,\frac12$). Data and corresponding fits (see text and table~\ref{relaxationLaws}) are shown in frames a)-e) as open circles, solid and broken lines, respectively. Frame f) emphasises the very abrupt change in the relaxation behaviour between 292 and 291~K. In frame e), the magnetisation recovery is also shown for irradiating one of the quadrupolar satellites (filled squares, down-shifted vertically for clarity). Where error bars are omitted, the uncertainty is less than the size of the symbols. \label{recovery}}
\end{figure}
The simplest relaxation is described by
\begin{equation}
m(t) = m_\infty-Ae^{-2Wt}\label{SingleExp},
\end{equation}
where $2W=T_1^{-1}$ denotes the SLRR. The saturation magnetisation is $m_\infty$ and $A$ is a fit parameter. Eq.~\ref{SingleExp} is valid for a narrow $^{23}$Na NMR line which can be fully irradiated.
In this case a clear identification of the possible relaxation mechanisms, i.e., magnetic, quadrupolar or mixed is, however,  not possible. In the situation of a dominant static Zeeman interaction (${\cal H}_Z$), a small quadrupolar perturbation (${\cal H}_Q$) and assuming purely magnetic fluctuations with a time scale $\tau_c\ll\nu_L^{-1}$ ($\nu_{\rm L}$: Larmor frequency), the magnetisation recovery, after a short-pulse irradiation of a single transition, is expected to follow
\begin{eqnarray}
m_{-\frac12}(t) &=& m_\infty-A\left(e^{-2Wt}+9e^{-12Wt}\right)\label{MagMultiExpCL}\\
m_{-\frac32, \frac12}(t) &=& m_\infty-A\left(e^{-2Wt}+5e^{-6Wt}+4e^{-12Wt}\right), \label{MagMultiExpW}\qquad
\end{eqnarray}
where $m_i(t)$ denotes the magnetisation recovery curve after a transition $i\leftrightarrow j$, with $i\in\{-I...I-1\}$ as the state with lower energy.
A purely quadrupolar relaxation mechanism, on the other hand, leads to
\begin{eqnarray}
m_{-\frac12}(t) &=& m_\infty-A\left(e^{-2W_1t}+e^{-2W_2t}\right)\label{QuadMultiExpCL}\\
m_{-\frac32, \frac12}(t) &=& m_\infty-A\left(e^{-2W_1t}+e^{-2(W_1+W_2)t}\right).\label{QuadMultiExpW}
\end{eqnarray}
Here, $W_1$ is related to EFG fluctuations in the $zx$ and $zy$ planes, leading to $\pm\frac12\leftrightarrow\pm\frac32$ nuclear transitions, while $W_2$ reflects the fluctuating EFG components in the $xy$ plane, allowing for transitions of the type $\mp\frac32\leftrightarrow\pm\frac12$~\cite{Bonera1970}. In the present case, $B_0\parallel c\parallel z$.
Figure~\ref{recovery} shows recovery data at selected temperatures together with respective fits (see also table~\ref{relaxationLaws}). Eqs.~\ref{MagMultiExpCL} and~\ref{QuadMultiExpCL} apply for the recovery after irradiation of the central line, eqs.~\ref{MagMultiExpW} and~\ref{QuadMultiExpW} are valid in the case when the satellites are irradiated.
\begin{table}
\centering\begin{tabular}{cccc}
\hline\hline
$T$  & ~Regime~ & ~Fig.~\ref{recovery}~ & ~SLRR~  \\
(K) & (A-D) & panels & eqn. \\
\hline
$<230$~K & A,B & d+e & (\ref{MagMultiExpCL}), (\ref{MagMultiExpW})\\
$~270-291$~K~ & C & c & (\ref{SingleExp})\\
$\geq292$~K & D & a+b & ~(\ref{QuadMultiExpCL}), (\ref{QuadMultiExpW})~\\
\hline\hline
\end{tabular}
\caption{Summary of differences in the relaxation behaviour in different temperature intervals. \label{relaxationLaws}}
\end{table}

The most obvious result, the drastic change in the central line relaxation behaviour between 292 and 291~K, is shown in fig~\ref{recovery}f. It indicates the first order transition that was previously reported, e.g., in ref.~\cite{Roger2007}, based on results from diffraction and resistivity experiments. Next, we note that the data in fig.~\ref{recovery}a and b are very well reproduced by using eq.~\ref{QuadMultiExpCL} (solid line), implying 2 different relaxation channels~\cite{Villa1980}. In the same temperature regime, the relaxation related to the satellites is very well described by using eq.~\ref{QuadMultiExpW}. In this way an unambiguous evaluation of the parameters $W_1$ and $W_2$ and their temperature dependences may be achieved.
The broken line in fig.~\ref{recovery}b demonstrates the inadequacy of the fit if the validity of eq.~\ref{SingleExp} is assumed. While the data in fig.~\ref{recovery}c is still best fit with eq.~4, recovery curves recorded at $T<290$~K are adequately approximated by using eq.~\ref{SingleExp}. The data at 200~K and below is best fit by using eq.~\ref{MagMultiExpCL}. An example of the satellite relaxation is shown in panel e) as solid squares. The corresponding fit employs eq.~\ref{MagMultiExpW}, with the same value for $W$ as that used in eq.~\ref{MagMultiExpCL} for fitting the central line recovery.

\begin{figure}
\includegraphics[width=\columnwidth]{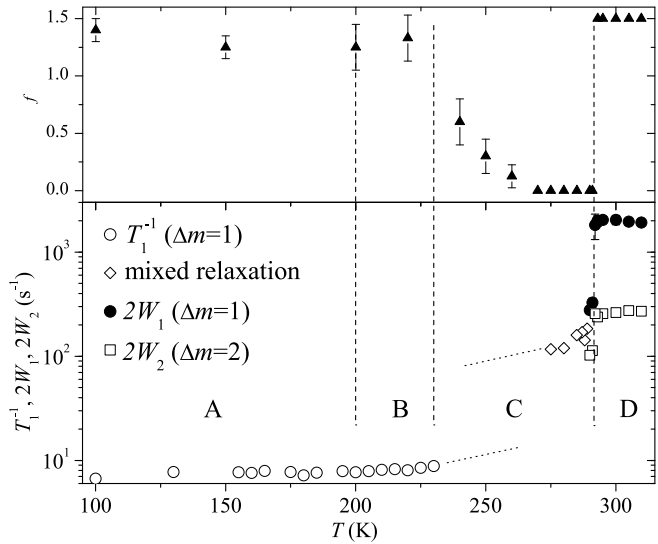}
\caption{The upper panel contains the relative intensity $f$ of the satellites with respect to the central line. In the lower panel, $^{23}$Na NMR spin-lattice relaxation rates of Na$_{0.8}$CoO$_2$ between 100 and 310~K are shown. The different regimes (A-D) are explained in the text. \label{T1T2}}
\end{figure}
\begin{figure}
\includegraphics[width=\columnwidth]{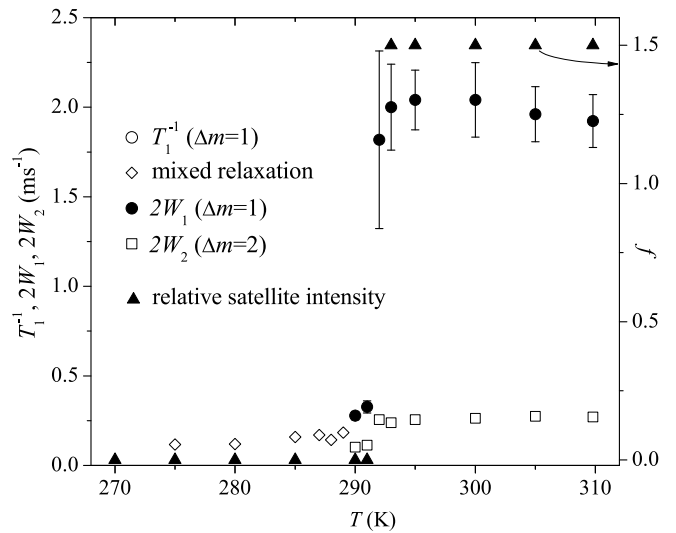}
\caption{Spin-lattice relaxation rates and relative satellite intensity (solid triangles) on an expanded temperature scale around the transition.\label{firstorder}}
\end{figure}

A compilation of the spin-lattice relaxation features after irradiation of the central line, captured in the temperature dependences of the characteristic parameters of the SLR channels, is presented in fig.~\ref{T1T2}. As will be outlined below, the entire $^{23}$Na NMR response is different in different parts (A-D) of the covered temperature range. Corresponding information is also contained in table~\ref{relaxationLaws}.
In fig.~\ref{T1T2} we note an almost temperature independent SLRR in region A which, as outlined above, is entirely due to magnetic fluctuations. The rate increases slowly with increasing temperature above 200~K. Above 230~K, the recovery data cannot satisfactorily be approximated by using eq.~\ref{MagMultiExpCL}, presumably because of a growing influence of quadrupolar relaxation and the later discussed cause of the concomitant reduction of the satellite intensity (see upper panel of fig.~\ref{T1T2}). Above 270~K, the now much faster relaxation can be well described by a single exponential using eq.~\ref{SingleExp}, but a separation into different channels is not possible. The situation is more transparent at 290~K and above. The relaxation is now dominated by fluctuations of the electric field gradient, leading to the behaviour captured in eq.~\ref{QuadMultiExpCL}. The transition is reflected in discontinuous enhancements of both $W_1$ and $W_2$ upon increasing temperature.
The upper panel of fig.~\ref{T1T2} displays the intensity ratio $f$ between the quadrupolar satellites and the central line. Above the transition, the theoretically expected value of 1.5, confirming that the total spectrum of the monitored Na nuclei is captured, is well established~\cite{Carter1977}. The abrupt quenching of the satellite signal at the transition is followed by a smooth recovery of the satellites with decreasing temperature in range C, starting at approximately 270~K. A trend to saturation of $f$ to an average value of 1.3 is noted in ranges~B and~A. Fig.~\ref{firstorder} emphasises these features between 270 and 310~K.

The interpretation of the NMR data set requires to consider various characteristic time scales that are related to the experimental method and the investigated material~\cite{Rigamonti1986, Suter1998}. Our experiment introduces the Larmor frequency $\nu_{\rm L} = \tau_{\rm L}^{-1}$ of the $^{23}$Na nuclei in the external field $B_0$. The duration of the present NMR experiments, $\tau_{\rm NMR}$,  covers the range from 20 to 1000~$\mu$s. Next we consider $\tau_h$, a hopping time related to the motion of Na ions and $\tau_c^{-1}$, the cut-off frequency limiting the range of the dynamic magnetic susceptibility $\chi(\omega)$ related to the localised moments on the Co sites. As we outline below, our data imply a significant reduction of $\tau_h$ with increasing temperature.

Considering the spectra in regime A shown in fig.~\ref{Spectra}a, the pattern of the different inequivalent Na-ion sites can be regarded as static and the NMR relaxation is dominated by the magnetic fluctuations of the Co moments. We argue that the electric field gradient acting on the Na-nuclei is mainly due to the Co-O sheets and therefore very similar for almost all of them, as reflected in the rather well defined satellite-bands. Apparently, the Na ions change sites only occasionally and hence $\tau_h$ is long compared to all NMR-relevant time scales~\footnote{In ref.~\cite{Schulze2008}, a characteristic time scale of approximately 2500~s was found for the Na ion rearrangement/relaxation at 200~K, i.e., 6-8 orders of magnitude longer than $\tau_{\rm NMR}$.}. The quality of the fits, using eq.~\ref{MagMultiExpCL} as described above, confirms that $\nu_{\rm L}\tau_c\ll1$. If $x<1$ ($x=0.8$ in our case), some thermally induced diffusion of Na ions may be expected. The increase of the SLRR in region B is compatible with this conjecture. In regime C, the increasing mobility of Na leads to a less well defined relaxation behaviour but at the upper end of this region, electric field gradient fluctuations due to Na motion dominate the relaxation features. Because of the gradual reduction of $\tau_h$ to below $\tau_{\rm NMR}$, each Na ion can occupy several different inequivalent positions during $\tau_{\rm NMR}$ and the resulting time-averaged electric field gradients are spread over a large range of values. In this regime only the Na-ions which do not change their position during the single NMR experiment contribute to the quadrupolar satellites. Consequently they loose intensity and disappear at approximately 270~K. The relaxation related to the single remaining line is described very well by eq.~\ref{SingleExp} between 270~K and approximately 290~K. The abrupt reappearance of the satellites and the narrow central line at $T\geq292$~K indicate that, on a time-average of $\tau_{\rm NMR}$, all Na nuclei experience the same local magnetic field and, considering the narrow satellites, nearly the same electric field gradient, respectively. This is compatible with very short times $\tau_h$ and $\tau_c$. We argue that this happens because the Na motion suddenly becomes so fast that the time-average of the electric field gradient is essentially the same for each Na-ion. Hysteresis effects in both the satellite intensity and the SLRR at the transition occur within less than 1~K. Because the $T_2^{-1}(T)$ data are not really important for the presently considered issues, we refrain from their presentation and detailed discussion. We only note that for $T\geq292$~K, $T_2^{-1}$ is of the order of the natural line width of the central line, namely of the order of 3~kHz.

If we consider the details in our NMR data across the transition, i.e., no significant change in the narrow signal of the central line, the abrupt reappearance of narrowed satellites and the discontinuous change in relaxation features (see fig.~\ref{firstorder}), it seems difficult to identify the transition to be of common structural type~\cite{Roger2007}. In particular, these data are not compatible with any structural change from  one static arrangement of inequivalent Na sites to another, a possible scenario if it is kept in mind that $x=0.8$.
Instead we interprete these observations to suggest that for $T\geq292$~K the Na layers adopt a 2D-liquid state, in the sense that the motion of the Na ions prevents the formation of a static structure with a well defined allocation of sites. Indeed the $^{23}$Na NMR response here resembles that previously observed in superionic conductors with planar Na layers~\cite{Villa1980}.

The reappearance of the quadrupolar satellites in this liquid state may seem counter intuitive. On the contrary, it is even expected if a 2D-type Na$^+$ liquid between negatively charged Co-O layers is considered. Also in this liquid phase, the Co-O layers are the main source of the electric field gradient, leading to the fact that the quadrupolar frequencies $\nu_Q$ above 292~K and below 270~K are almost identical. This again is difficult to reconcile with only a reordering of the Na ions at the transition.

The data obtained in this work is also helpful for the interpretation of previously reported puzzling results concerning $^{23}$Na $T_1^{-1}(T)$ measured on powdered samples of Na$_{0.7}$CoO$_2$~\cite{Gavilano2004a}. With hindsight it may be concluded that strong Na diffusion and the transition documented here, are also features of this material.


\begin{acknowledgments}
We thank  H. R. Aeschbach, R. Hindshaw, J. Kanter, R. Monnier, T. Schulze and V. Wittwer for their help in the preparation of the experiments, sample characterisation and for valuable discussions. This work was, in part, financially supported by MaNEP, an NCCR of the Swiss National Science Foundation (SNF).
\end{acknowledgments}

\end{document}